\documentclass[12pt]{article}
\advance\hoffset by -1.1truecm
\advance\voffset by -1.0truecm
\begin{document}

\title{\Large On the codon assignment of chain termination signals
 and the minimization of the effects of frameshift mutations}

\author{Jean-Luc Jestin\thanks{email: jlj@mrc-lmb.cam.ac.uk
}\\
\small Centre for Protein Engineering and Laboratory of Molecular Biology\\
\small Medical Research Council,  Cambridge CB2 2QH,  UK\\
\rm Achim Kempf\thanks{email: a.kempf@amtp.cam.ac.uk}\\
\small Corpus Christi College in the University of Cambridge\\
\small Cambridge CB2 1RH, UK \rm}

\date{}
\maketitle

\begin{abstract} It has been suggested that
the minimization of the probability for lethal mutations 
is a major constraint shaping the genetic code \cite{1}.
Indeed, the genetic code has been found
highly protective against transitions \cite{2}. Here, we show that data on
polymerase-induced frameshifts provide a rationale for the codon assignment
 of chain termination signals (CTS).
\end{abstract}

We work on the assumption that
the mutational spectra of in vitro polymerization \cite{3}-\cite{11}
for DNA-polymerases belonging to families 
found in at least two of the three living 
kingdoms \cite{12} are relevant with respect to primordial polymerases.
We will here not take into account DNA
polymerases beta, which are family X DNA polymerases exclusively found among
eukaryotes so far \cite{13}, and HIV-reverse transcriptases, 
which are thought to
be active as dimers \cite{14} and which emerged very `late' 
in evolution \cite{15}.

As it is believed that RNA preceded 
DNA in evolution, data for RNA-replicases
would be more adapted but are not
 available. Recent evidence shows however 
that DNA- and RNA-replicases are very 
closely related \cite{16}: a single tyrosine to
phenylalanine substitution changes
 DNA-replicases into RNA-replicases \cite{17}; E.coli
DNA-polymerase I is also an accurate 
RNA-dependent DNA-polymerase \cite{18}. A single
mutation confers to the MMLV-reverse 
transcriptase the ability to replicate
 RNA \cite{19}.

Polymerase-induced mutations are mainly
 substitutions and frameshifts \cite{3}-\cite{9}. The
frameshifts' error-rate is about half
 the substitutions' error-rate for the
Klenow-polymerase domain \cite{3}, which 
has no nuclease domain as can be assumed for a
primordial polymerase. Frameshifts result
 mostly from the addition or deletion of
one base \cite{3}-\cite{7}. Frameshifts are highly 
deleterious as they prevent translation in the
correct reading frame of the codons 
downstream the mutation. They happen in
directly repeated and palindromic
 sequences \cite{10} (where the assumption of
polymerase-error tolerance can be 
shown to be consistent) and in
non-reiterated runs, where single-base 
deletions occur more frequently than
single-base additions \cite{3,4},\cite{6},\cite{9}.
Additions will therefore be neglected here. For
polymerases with or without nuclease
 domains we noticed no significant
differences  in the consensus sequence 
for single-base deletion sites in
non-reiterated runs. It has been defined as YR \cite{3}, TTR \cite{11},
 YTG \cite{8}, TR \cite{10}, and refined
here from the current data \cite{3}-\cite{11} 
as YTRV (V= C, A or G \cite{3}).

If the genetic code has been optimized for 
frameshift tolerance, then it should 
allow to code for most amino acid
sequences without using YTRV sequences,
 or the TRV,
YTR and NYT potential deletion site
 codons (PDSC) nor using their
reverse-com\-ple\-men\-tary (rc)
 sequences which are also expected to yield deletions
during replication (Fig. 1): 

If the base T is the first base of a codon and in
case the previous codon has a pyrimidine 
as third codon base, then the amino acid
should be encoded without using the 
six codons TRV; if the base T is the second
base of a codon and in case the first
 base of the following codon is C, A or G,
then the amino acid should be encoded 
without using the codons YTR; if the 
base T is the third base of a codon 
and in case the following amino
 acid has a RVN-type codon, then the
 amino acid should be encoded
 without using the eight codons NYT.

The most deleterious codons are TAA 
and TAG and their rc-sequences TTA and 
CTA
that are both PDSC and rc-PDSC. Deletions 
at codons encoding amino acids are
likely to yield non-functional proteins, as all downstream codons are not
translated.  However,
 deletions at codons encoding CTS should result in addition of
peptides to the proteins' carboxy-termini,
 thereby likely providing functional
proteins. \it The least deleterious effects are therefore 
obtained by assigning the most 
deleterious codons to CTS and not to amino acids. \rm 
Frameshift tolerance may
therefore have been the major constraint in the codon assignment of CTS.
The codons TTA and CTA encode leucine having the highest, six-fold degeneracy.
Frameshift tolerance may therefore be one of the constraints imposing a high
degeneracy to these amino acids.

We point out, however, that substitution tolerance and frameshift
mutation tolerance are to be considered as competing
constraints on the selection of an optimal genetic code:
substitution tolerance favors a code
 in which an amino acid is encoded by triplets
differing only by single-base mutations.
 On the other hand, given that single-base
deletions in non-reiterated runs
 occur mostly on a specific template sequence,
(YTRV), tolerance of these frameshift 
mutations favors a code in which amino 
acids are encoded by triplets differing 
strongly from another, so that amino 
acid sequences are more likely to be
 able to be coded for without using 
the YTRV sequences. Our argument is based on the 
consideration that, for the consensus sequences,
the ratio between single-base deletions and base substitutions is
much greater than in other sequences, so that, 
for the assignment of specific codons, 
frameshift mutation tolerance should be a stronger constraint 
than substitution tolerance.

In conclusion, these results provide an insight 
into the constraints yielding the
genetic code's fixation and suggest that
 the codon assignment of CTS may be
contemporary with the emergence of polymerases being enzymes rather than
ribozymes. Polymerase-error tolerance arguments similar to the one 
presented here may be useful in the investigation 
of alternative terrestrial
or exobiological genetic codes and possibly 
also for the engineering of new 
genetic codes.
\medskip \newline
Tables available from autors via email and mail.


\begin{thebibliography}{**}
\small
\bibitem{1} Sonneborn T.M., in Evolving genes
 and proteins (ed. Bryson V. {\&} Vogel H.J.) 377-
397, (Academic Press, New York, 1965).
\bibitem{2} Goldberg A.L., Wittes R.E., Science, 
153, 420-424, (1966).
\bibitem{3} Bebenek K., Joyce C.M., Fitzgerald M.P.,
 Kunkel T.A., J. Biol. Chem., 265, 
13878-13887, (1990).
\bibitem{4} Bell J.B., Eckert K.A., Joyce C.M., 
Kunkel T.A., J. Biol. Chem., 272, 7345-
7351, (1997).
\bibitem{5} Minnick D.T., Astatke M., Joyce C.M., 
Kunkel T.A., J. Biol. Chem., 271, 24954-
24961, (1996).
\bibitem{6} Kunkel T.A., Patel S.S., Johnson K.A., 
Proc. Natl. Acad. Sci. USA, 91, 6830-
6834, (1994).
\bibitem{7} Kunkel T.A., J. Biol. Chem., 260,
 12866-12874, (1985).
\bibitem{8} Papanicolaou C., Ripley L.S., J. Mol.
 Biol., 207, 335-353, (1989).
\bibitem{9} Cai H., Yu H., McEntee K., Kunkel T.A.,
 Goodman M.F., J. Biol. Chem., 270, 
15327-15335, (1995).
\bibitem{10} Wang F.J., Ripley L.S., Genetics, 136, 
709-719, (1994).
\bibitem{11} De Boer J.G., Ripley L.S., Genetics, 
118, 181-191, (1988).
\bibitem{12} Braithwaite D.K., Ito J., Nucl. Acid. 
Res., 21, 787-802, (1993).
\bibitem{13} Bork P., Ouzounis C., Sander C., Scharf M., 
Schneider R., Sonnhammer E., Prot.
Sci., 1, 1677-1690, (1992).
\bibitem{14} Restle T., M=FCller B., Goody R.S., J. 
Biol. Chem., 265, 8986-8988, (1990).
\bibitem{15} Eigen M., Nieselt-Struwe K., AIDS, 4, 
S85-S93, (1990).
\bibitem{16} Joyce C.M., Proc. Natl. Acad. Sci.
 USA,  94, 1619-1622, (1997).
\bibitem{17} Sousa R., Padilla R., EMBO J., 14,
 4609-4621, (1995).
\bibitem{18} Ricchetti M., Buc H., EMBO J., 12, 
387-396, (1993).
\bibitem{19} Gao G., Orlova M., Georgiadis M.M., 
Hendrickson W.A., Goff S.P., Proc. Natl.
Acad. Sci. USA, 94, 407-411, (1997).
\end{thebibliography}
\end{document}